\documentclass[twocolumn,showpacs,prb,amsmath,amssymb,floatfix]{revtex4}

\usepackage{graphicx}
\usepackage{dcolumn}
\usepackage{bm}
\usepackage{latexsym}
\usepackage[usenames]{color}

\makeatletter
\def\lsim{\mathrel{\mathpalette\gls@align<}}
\def\gsim{\mathrel{\mathpalette\gls@align>}}
\def\gls@align#1#2{\lower.6ex\vbox
	{\baselineskip\z@skip\lineskip\z@\ialign
		{$\m@th#1\hfill##$\crcr#2\crcr\sim\crcr}}}
\makeatother
\def\be{\begin{equation}}
\def\ee{\end{equation}}
\def\bea{\begin{eqnarray}}
\def\eea{\end{eqnarray}}

\makeatletter
\def\leqq{\mathrel{\mathpalette\gle@align<}}
\def\geqq{\mathrel{\mathpalette\gle@align>}}
\def\gle@align#1#2{\lower.6ex\vbox{%
\baselineskip\z@skip\lineskip\z@\ialign{$\m@th#1\hfill##$\crcr#2\crcr=\crcr}}}
\makeatother
\newcommand{\vek}[1]{\mbox{\boldmath${#1}$}}

\begin{document}
\preprint{Step Repulsion Mediates Wandering on Si(001) Vicinal Face}
\title{Step Repulsion Mediates Wandering on a Si(001) Vicinal Face}

\author{Masahide Sato$^{1,}$}\email{sato@cs.s.kanazawa-u.ac.jp}
\author{Makio Uwaha$^2$}
\author{Yukio Saito$^3$}
\author{Yukio Hirose$^1$}
\affiliation{
	$^1$Department of Computational Science, Kanazawa University,
		Kakuma-cho, Kanazawa 920-1192, Japan\\
	$^2$Department of Physics, Nagoya University,
		Furo-cho, Chikusa-ku, Nagoya 464-8602, Japan \\
	$^3$Department of Physics, Keio University,
		3-14-1 Hiyoshi, Kohoku-ku, Yokohama 223-8522, Japan
}
\date{\today}

\begin{abstract}
With a Si(001) vicinal surface in mind, we study step wandering
instability on a vicinal surface with an anisotropic surface diffusion
whose orientation dependence alternates on each consecutive terrace.
In a conserved system step wandering takes place with step-up adatom drift. 
Repulsive interaction between steps is found indispensable 
for the instability.
Monte Carlo simulation with a strong repulsive step interaction
confirms the result of linear stability analysis,
and further shows that in-phase step wandering produces straight grooves.
Grooves widen as their amplitudes increase 
in proportion to the square root of time.
\end{abstract}


\maketitle

On a vicinal surface of a crystal, steps undergo 
two types of dynamical instabilities: wandering and bunching.
Step wandering is the instability for step deformation along the step,
and step bunching is the one for the inter-step distance~\cite{rev1}.
Both instabilities are caused by some asymmetry in the 
surface diffusion field.
There are many effects which cause the asymmetry, and
the drift flow of adsorbed atoms (adatoms) is one of them.
On Si(111) and Si(001) vicinal faces,
 a direct electric current induces drift of adatoms
and the instabilities have been observed under its application~\cite{rev2}.

\begin{figure}[btph]
       \centerline{
       \includegraphics[width=0.95\linewidth]{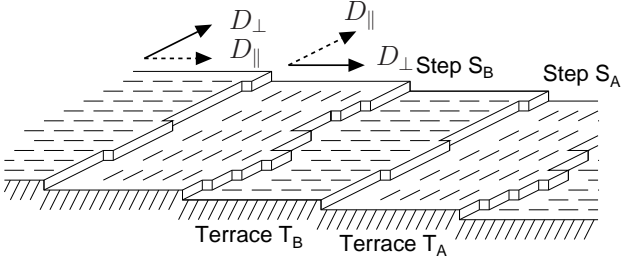}
       }
       \caption{A Si(001) vicinal face. 
       Short lines on terraces represent dimers.
       }\label{fig:vicinal}
\end{figure}

Si(001) surface is reconstructed and forms rows of dimerized atoms
arranged in a $2 \times 1$ unit cell (Figure~\ref{fig:vicinal}).
On the reconstructed surface, the adatom surface diffusion is anisotropic
such that it takes place more easily in parallel to the dimer rows 
than in perpendicular.
On a vicinal face terraces of different heights are bounded by steps.
On consecutive terraces the dimer orientation alternates, and we call the 
$1 \times 2$ terrace T$_A$ and the $2 \times 1$ terrace T$_B$.

Due to the alternation of the orientation of fast surface diffusion on
different terraces, conditions of the step instabilities for a Si(001)
vicinal face differ from those for a Si(111) vicinal face.
Experimentally, bunching is observed on a (001) vicinal face 
with a finite current irrespective of its
direction~\cite{Kahara-y89jjap,Litvin-kl91,Latyshev-la98ass},
and the step wandering with the step-up current~\cite{Nielsen-pp01ss}.
Since the drift is believed to be parallel to the 
current~\cite{Metois-hp99ss}, the drift direction to cause the step
wandering is opposite to that on a Si(111) vicinal face.

Theoretically, step bunching on a Si(001) vicinal face
is studied by a one-dimensional step flow 
model~\cite{Stoyanov-90jjap,Natori-fy92jjap,Natori-ff92ass}
and by Monte Carlo simulations~\cite{Sato-us02jcg}.
When the alternation of anisotropic surface diffusion
is taken into account, the step bunching instability is found
irrespective of the drift direction, in agreement with the 
experiments~\cite{Kahara-y89jjap,Litvin-kl91,Latyshev-la98ass}.
On the contrary, there is no theoretical study on the step wandering
on a Si(001) vicinal face so far, which we undertake in this paper.

Atoms detached from steps migrate on terraces and attach to some steps.
For the wandering instability, the adatom drift is necessary as well
as the diffusive motion. Evaporation and impingement are omitted.
Steps are running parallel to the $x$-direction on average,
and the positive $y$-direction is chosen in the step-down direction.
The drift is assumed in the $y$-direction.
With the anisotropy of the diffusion coefficient,
the diffusion equation of adatom density $c(x,y,t)$ is expressed as
       \begin{equation}
       \partial_t c =
       D_{x} \partial_{x}^{2 } c +D_{y}\partial_{y}^{2 }c
       -f D_y\partial_{y}c,
       \label{eq:diffusion}
       \end{equation}
where $\partial_t $, $\partial_x $
 and $\partial_y$ represent
the partial derivatives with respect to the subscripts,
$D_{x}$ is the diffusion coefficient in the $x$-direction,
$D_{y}$  in the $y$-direction, and $f$ ($=F/k_{\rm B}T$) 
the force to induce the drift devided by the temperature.
The meaning of $D_{x}$ and $D_{y}$ depends on which terrace we are
discussing, on T$_A$ or T$_B$, and we come to this point later.
For simplicity, we assume that the step kinetics is fast enough 
that the adatom density  attains its local equilibrium value at 
each step:
$\left. c \right|_\pm 
=c_{eq}^0 [ 1+ \Omega (\tilde{\beta}\kappa + \partial_y  \zeta )/k_{\rm B}T]$. 
Here  $+(-)$ indicates the lower (upper) side of the step,
$c_{\rm eq}^0$ is the equilibrium adatom density of a free straight step,
$\Omega$ is the atomic area, $\tilde{\beta}$  the step stiffness,
$\kappa$  the curvature of the  step, and $\zeta$ the step-step interaction
potential. We assume that  $\zeta$ is a function of the step distance $l$
in the $y$-direction as 
$  \zeta  = A (l_+^{-2}+l_-^{-2}) $ with a positive constant $A$,
corresponding to step repulsion.
There is a more detailed model where $\zeta$ is expressed by an
integration of the force dipole along the step~\cite{Paulin-gpm01prl}.
But in the linear analysis the complication is shown to be incorporated 
into the renormalization of the stiffness. Therefore, we use the simple
form for $\zeta$  here.

By solving eq.~(\ref{eq:diffusion}) with the boundary conditions 
in the quasi-static approximation ($\partial_t c  =0$),
the adatom density $c$ and then the adatom current $\mbox{\boldmath $j$}$ 
are determined. The step velocity is given by 
$V= \Omega \mbox{\boldmath $\hat{n}$}\cdot \left( \left. \mbox{\boldmath $j$}
\right|_{-} - \left. \mbox{\boldmath $j$} \right|_{+}\right),$
where $\mbox{\boldmath $\hat{n}$}$ is the normal vector 
in the step-down direction.

Due to the different orientation of the dimer rows on T$_A$ and T$_B$,
a set of diffusion coefficients $(D_x, D_y) $ corresponds different
combinations; $(D_x, D_y) = (D_\parallel, D_\perp)$ on T$_A$
and $(D_x, D_y) = (D_\perp, D_\parallel)$ on T$_B$, and
$D_\perp$ is larger than $D_{\parallel}$.
Since the step S$_B$ is rougher than S$_A$ on a Si(001) vicinal face,
step parameters are different in general for the two types of steps, 
S$_A$ and S$_B$, but for  simplicity, we neglect these differences.

On a flat vicinal face where parallel steps are arranged equidistantly, 
the adatom concentration is homogeneous as $c_{\rm eq}^0$.
The step velocities $V_A$ and $V_B$ of the steps 
S$_A$ and S$_B$ are 
calculated as $V_A=-V_B=-\Omega f(D_{\perp} -D_{\parallel})c_{\rm eq}^0.$
Since the step velocities $V_A$ and $V_B$ 
are nonvanishing and opposite for a finite drift ($F \neq 0$),
the flat vicinal face is unstable against step pairing.
Without the repulsive step-step interaction,
the adatom concentration is homogeneous as $c_{\rm eq}^0$
irrespective of the step distance. 
The steps move with the velocities $V_A$ and $V_B$ given above, and by
coalescence the surface is covered by one type of terrace; for instance,
T$_A$ for positive $F$ since $D_{\perp} > D_{\parallel}$.

If the repulsive interaction is granted,
the difference of terrace widths $l_{\rm A}$ and $l_{\rm B}$ of
terraces T$_A$ and T$_B$ causes the difference of the equilibrium 
adatom densities $c_{\rm A}$ and $c_{\rm B}$ at steps S$_A$ and S$_B$. 
Then the steady state with vanishing step velocities can be established 
even with the drift as
        \begin{equation}
        \frac{D_{\perp}(c_{\rm A}{\rm e}^{fl_{\rm A}}-c_{\rm B})
        }{{\rm e}^{fl_{\rm A} } -1} 
        =\frac{D_{\parallel}
        (c_{\rm B}{\rm e}^{fl_{\rm B}  }-c_{\rm A})
        }{{\rm e}^{fl_{\rm B} } -1}.
        \label{eq:lalb}
        \end{equation}

Since the average terrace width of the vicinal face is fixed to $l$,
wide and narrow terraces appear alternately as 
$l_{\rm A} = l + \delta l/2 $ and $ l_{\rm B} = l-\delta l/2$,
and the  equilibrium densities also alternate as
$c_\mathrm{A} =c_\mathrm{eq}^0 - \Delta c/2$ 
and $c_\mathrm{B} =c_\mathrm{eq}^0 + \Delta c/2$, respectively.
For a small drift $fl$ and the strong repulsive interaction,
the density difference $\Delta c=c_B-c_A$ is approximated by
$\Delta c / c_\mathrm{eq}^0 \approx \sigma fl $
with $\sigma = (D_\perp-D_\parallel)/(D_\perp+ D_\parallel)$, 
and the terrace width difference is 
$\delta l /l \approx  k_BT \sigma fl^2 /  12  \Omega A $.
Without repulsive interaction $A=0$, the terrace width cannot remain finite.
Also the above equation indicates that the deviation of the terrace width 
$\delta l/ l$ is small under a very strong step repulsion.

We now study the stability of this steady state with alternating terrace
widths under the adatom drift. When the step repulsion is strong enough,
the perturbations to step positions $\delta y_\mathrm{A}(x,t)$
and  $\delta y_\mathrm{B}(x,t)$ of a wavenumber $q$ can be shown to merge
into the in-phase fluctuation of the same amplitude
$\delta y_\mathrm{A} (x,t)= \delta y_\mathrm{B} (x,t) 
=\delta y_q \mathrm{e}^{\omega_q t}   {\rm e}^{iq x}$.
With this deformation the terrace widths in the $y$ direction remain 
$l_{\rm A}$ and $l_{\rm B}$ as before. For small $q$, 
the amplification rate  is calculated to be
       \begin{equation}
       \omega_q = \alpha_2 q^2 -\alpha_4 q^4,
       \label{eq:omega1}
       \end{equation}
where $\alpha_2 = -\Omega (D_\perp -D_\parallel) \Delta c/2$ and 
$\alpha_4=\Omega^2 c_{eq}^0 
(D_\parallel l_\mathrm{A} + D_\perp l_\mathrm{B})
\tilde{\beta}/2k_\mathrm{B} T.$
Here we have assumed that 
the terrace widths are small as $q l_{\rm A, \rm B} \ll 1$.

When the drift is in the step-down direction($F>0$),
$l_{\rm A}$ is larger than $l_{\rm B}$,
and $c_{\rm B}$ is larger than $c_{\rm A}$.
Then the quadratic term in $q$ is negative
and the steady state is stable. 
When the drift is in the step-up direction($F<0$),
$c_{\rm A}$ is larger than $c_{\rm B}$, and
the quadratic term is positive. Then the steady state with straight steps
 is unstable and in-phase step wandering occurs. 
Thus the first term indicates an instability of the steady state by
changing its sign with the drift direction.
Note that the wandering is caused by the difference 
$\Delta c$ of equilibrium concentrations 
at the steps originated from the step repulsion.
Thus the step repulsion is indispensable for the establishment
of wandering instability on a Si(001) vicinal.
Also note that there is no critical value for the step-up drift.
If the drift changes from step-down to step-up, steps immediately
shows the wandering instability, because 
the Gibbs-Thomson effect gives only a quartic term,
higher than the destabilizing quadratic term.

For a strong repulsive interaction, 
the difference of the step distance $\delta l$ is small and 
the wavelength of the most unstable mode is given by
$ \lambda_\mathrm{max}  =2 \pi \sqrt{2 \alpha_4/\alpha_2}
=2 \pi \sigma^{-1} \sqrt{2 \Omega \tilde{\beta}/|F|}.$
The characteristic wavelength is inversely proportional 
to the square root of the external field $E$
which causes the drift:  $\lambda_\mathrm{max}\propto 1/\sqrt{E}$.

       \begin{figure*}[thbp]
       \centerline{
       \includegraphics[width=0.3\linewidth]{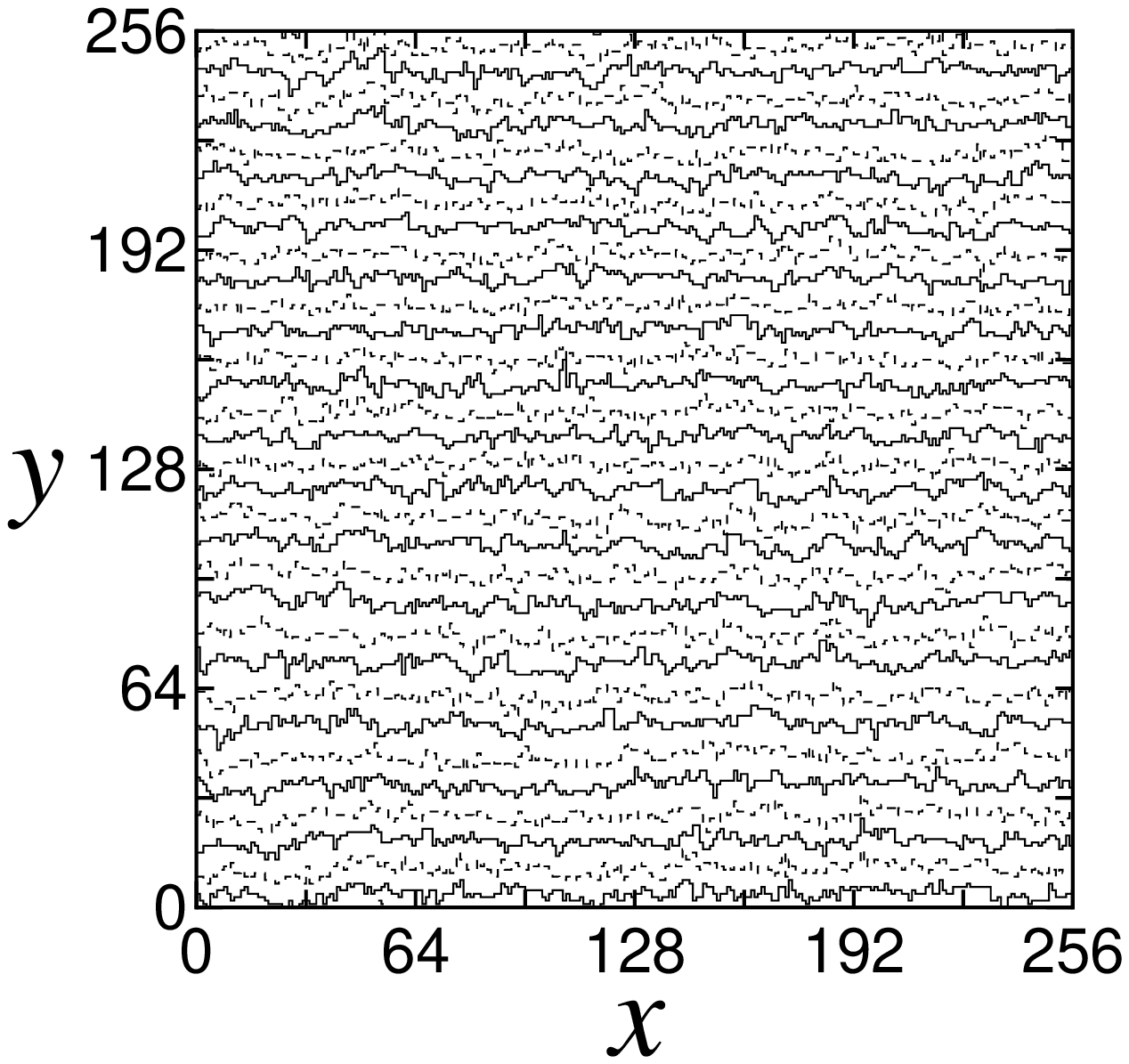}
       \includegraphics[width=0.3\linewidth]{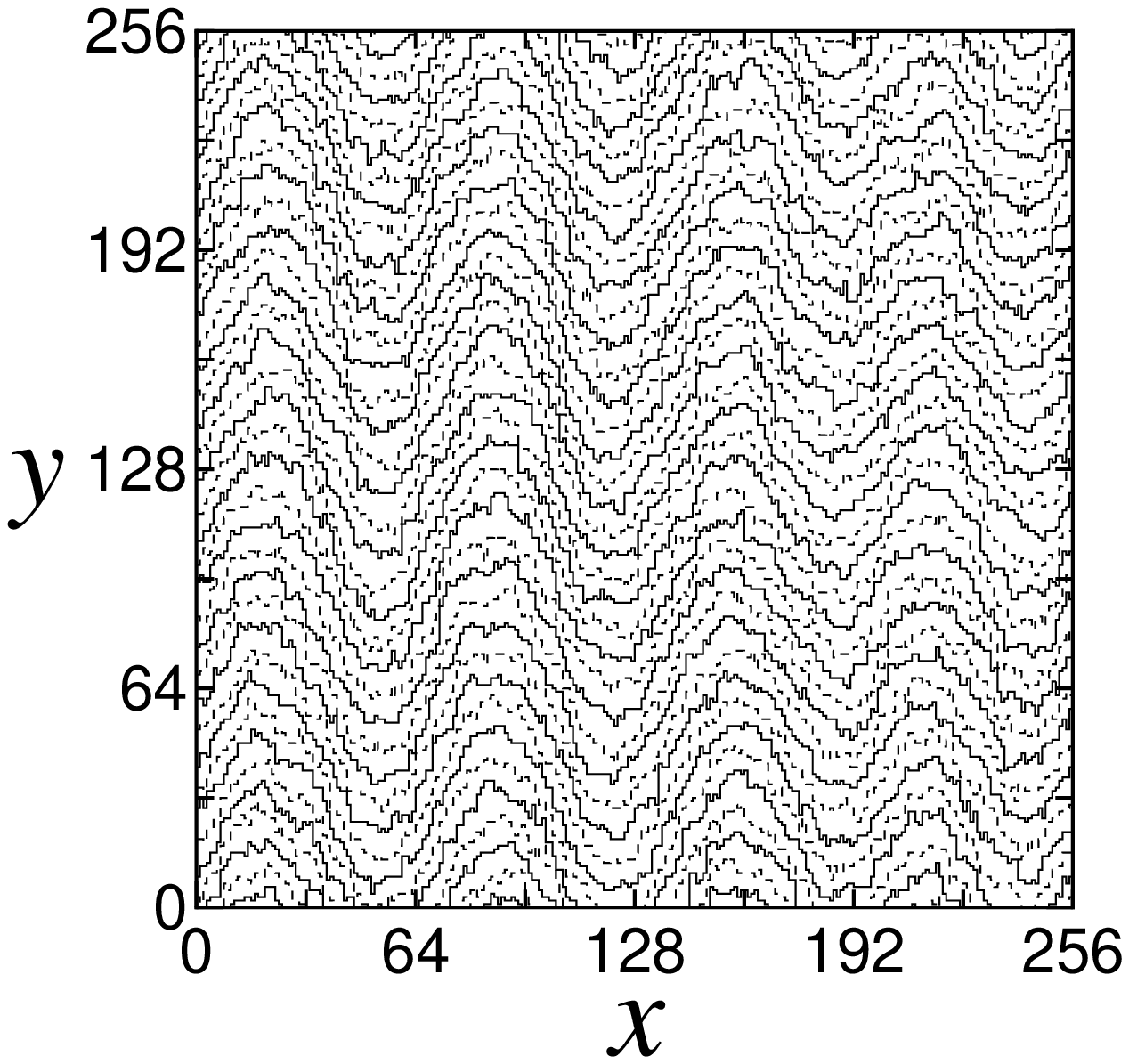}
       \includegraphics[width=0.3\linewidth]{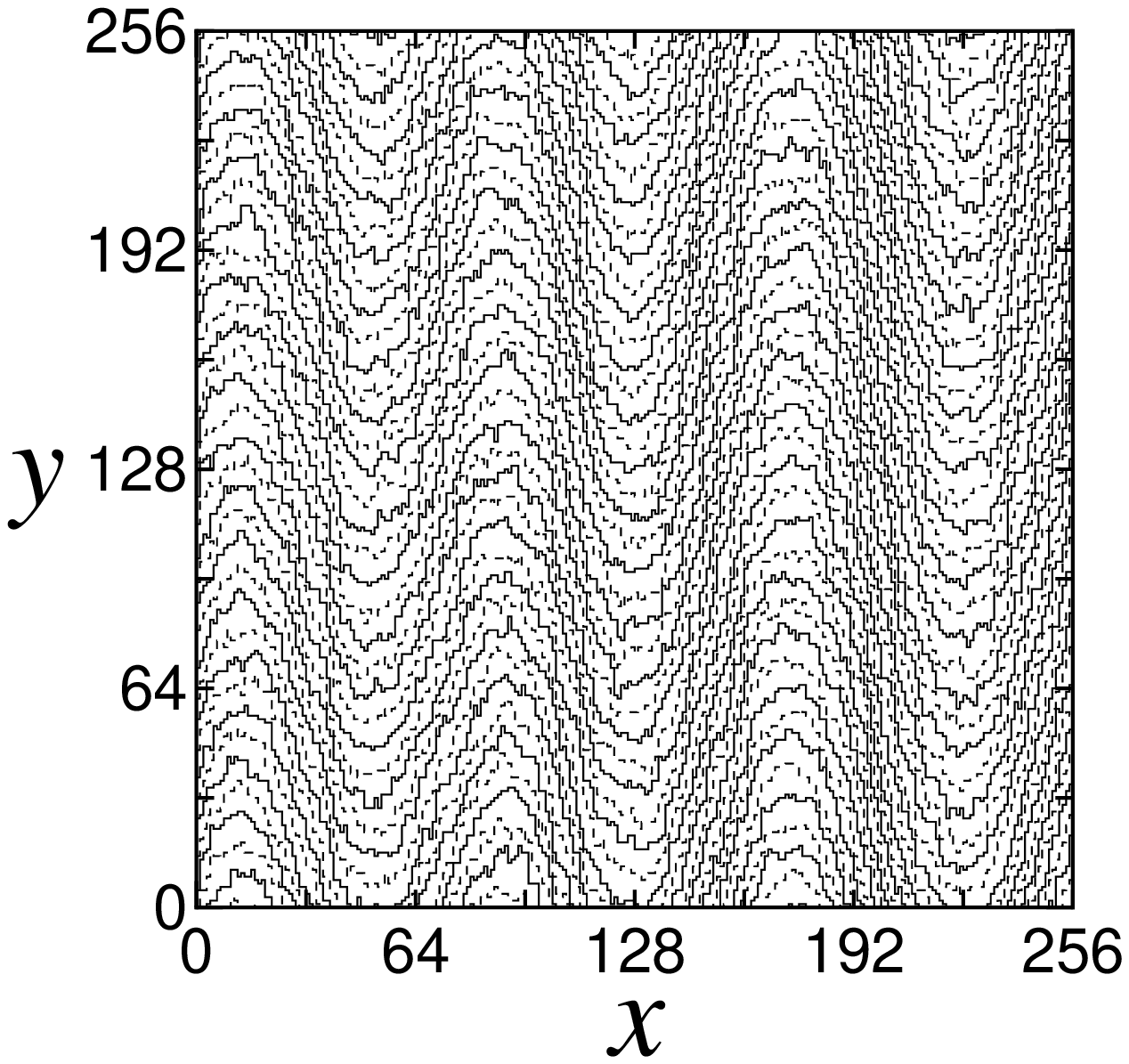}
       }
       \centerline{\large \sf (a)\hspace*{5.0cm}(b)\hspace*{5.0cm}(c)}
       \caption{
       Snapshots of the step wandering 
       (a) with step-down drift at $t \approx 5.0 \times 10^5$,
       (b) with step-up drift at $t \approx 2.5 \times 10^5$,
       and (c) with step-up drift at $t \approx 12.4 \times 10^5$.
       The number of steps is 32 and the system size is $256 \times 256$.
       }\label{fig:snapshot}
       \end{figure*} 

After the instability sets in, the step deformation 
amplifies,  and a nonlinear analysis is wanted.
Assuming an in-phase motion of steps due to the strong step repulsion, 
a heuristic argument is possible on the nonlinear evolution 
of the step wandering $\delta y(x,t) =\eta(x,t)$. 
This simultaneously reveals the physical origin of the wandering. 
Since the number of atoms is conserved, 
the in-phase step motion is controlled 
by the adatom current in the $x$ direction. 
The drift current has only the $y$ component so that only 
the diffusion current determines the evolution of step fluctuation. 

There are two contributions of diffusion current. 
One is the current  across the terrace induced by
 the difference of the equilibrium  densities at both ends of the terrace. 
When two bounding  steps are tilted from the $y$ direction with an angle 
$\theta=\tan^{-1}(\partial_x \eta)$, the gradient of the adatom density
in a terrace  T$_\mathrm{A}$ is 
$\vek{\nabla} c_A=(\Delta c/l_{A}\cos\theta)(-\sin \theta, \cos \theta)$.
The diffusion current is obtained as 
$\vek{j}^A=(D_\parallel \Delta c \tan \theta/l_A, 
-D_\perp \Delta c/l_A)$,
and the $x$ component of the total flux on T$_A$ is
$ J_x^A=j_x^Al_A=D_\parallel \Delta c \tan \theta $
(one obtains the same result by solving (\ref{eq:diffusion})).
Similarly the $x$ component of the flux on T$_B$ is 
$J_x^B=-D_\perp \Delta c \tan \theta$, and the average flux per step is
$J_x^{(1)}=(J_x^A+J_x^B)/2
=-(D_\perp-D_\parallel) \Delta c \,\partial_x \eta /2 $.
A steady state condition similar to eq.~(\ref{eq:lalb}) 
with tilted steps gives the difference of the equilibrium adatom
 density as  $\Delta c=\sigma fl c_{\rm eq}^0\,\cos^2\theta$, 
and the extra factor $\cos^2\theta$ appears. 
When $\Delta c<0$ ({\it i.e.} $F<0$), the flux is  an increasing function
of the slope and  an instability is expected.

The other is the diffusion current along the steps
due to the chemical potential change with curvature. 
Considering that the normal distance between steps for a tilted part 
is $l_{A,B}\cos\theta$, the flux (per step) passing through the two
terraces  induced by the chemical potential gradient is 
$ J^{(2)}_x = - 
(l_A\cos \theta D_{\parallel} +l_B\cos \theta D_{\perp})
c_\mathrm{eq}^0 \cos\theta \partial_x \mu/2k_\mathrm{B}T.$
The chemical potential $\mu$ is determined solely by the Gibbs-Thomson effect 
$\Omega \tilde{\beta} \kappa $ and independent of the step repulsion
in the present choise of interaction $\zeta$, since the 
terrace width in the $y$ direction is constant for the in-phase modulation. 
More general expression given by Paulin and co-workers~\cite{Paulin-gpm01prl} 
might modify the following result quantitatively but not qualitatively. 

With two contributions together, mass conservation leads to the following
nonlinear time evolution of the in-phase step deformation $\eta(x,t)$;
       \begin{eqnarray}
       \partial_t \eta
       &=&
       -\partial_x\left[ g^{-1} \left(
       \alpha_2  \partial_x \eta + \alpha_4 \partial_x
       \left(g^{-3/2}  \partial_x^2 \eta \right) \right)\right], 
       \label{eq:nonlinear}
       \end{eqnarray}
where  and $g = 1+  (\partial_x \eta) ^2=\sec^2\theta$.
$\alpha_2$ and $\alpha_4$ are those in eq.~(\ref{eq:omega1}).
With $\eta = \delta y_q e^{iqx + \omega t}$ 
the linear amplification rate $\omega_q$
is recovered from eq.(\ref{eq:nonlinear}).
Interestingly eq.(\ref{eq:nonlinear}) is the same as 
the nonlinear equation obtained for other conserved 
systems~\cite{Pierre-Louis-mskp98prl,Gillet-pm00epjp,%
Paulin-gpm01prl,Sato-ush2002prb,kato-ush} although the mechanism
looks very different.

The above theoretical analysis is now compared with the results of 
Monte Carlo simulations. For the simulation algorithm, 
consult Refs.\onlinecite{Sato-us02jcg,Sato-us2000prb,Saito-u94prb}.  
We study steps with an average distance $l=8$ 
on a square lattice system of size 256 $\times$256 or 512 $\times$ 128.
Length hereafter is measured in the unit of lattice constant $a$,
and time in the unit of $a^2/D_{\perp}$.
The parameters are so chosen as
the equilibrium adatom density $c_\mathrm{eq}^0=0.18$, 
the step stiffness $\tilde{\beta}/k_\mathrm{B}T =0.13$, 
$D_{\parallel} =0.5$ and $D_\perp =1.0$. 
There is no extra energy barrier for the over-step diffusion. 
Kinetic coefficient is large enough so that the local 
equilibrium condition  is valid.
The strength of repulsive potential $A/k_{\rm B}T =46$ is 
large enough to suppress step bunching in the following simulations.

In Figure~\ref{fig:snapshot} 
we show snapshots of the step wandering under various drift forces;
$f =  0.1$ for Fig.~\ref{fig:snapshot}(a) and -0.1 
for Fig.~\ref{fig:snapshot}(b) and (c).
Solid lines represent the step S$_A$ and dotted lines 
represent the step S$_B$.
We start the simulation with an equidistant train of straight steps.
With step-down drift in Fig.~\ref{fig:snapshot}(a)
(upward drift in the figure), steps remain straight.
With step-up drift in Fig.~\ref{fig:snapshot}(b) and (c)
step wandering occurs, in agreement with the linear stability analysis.
Because the wandering is in phase, grooves appear parallel to the $y$-axis.

Since $|f l|=0.8$ is not very small,
we have to use a general formula for the  wavelength 
of the most unstable mode, and it is obtained as
$\lambda_{\rm max} \approx 77$, in good agreement with the result 
$\lambda \approx 64$ in Fig.~\ref{fig:snapshot} (b). There, 
the wandering amplitude (the average step fluctuation width)
is $w \approx 14.5$. 
In a late stage shown in Figure~\ref{fig:snapshot}(c) 
the amplitude increases up to $w \approx 37.2$ 
when the wavelength of the grooves is about $85$.
Thus, the structure coarsens parallel as well as perpendicular to the steps.

Recently, Paulin and co-workers~\cite{Paulin-gpm01prl} 
studied the step wandering induced by the Ehrlich-Schwoebel effect
in a conserved system.
They found perpetual enhancement of the wandering amplitude 
as $w \sim t^{\beta} $ with $\beta \sim 1/2$,
 irrespective of the step repulsion, but the
coarsening of the wavelength of grooves took place only 
with a step repulsion.
Although we take account of the long-ranged step repulsion only 
in the $y$-direction but not in the $x$-direction as Paulin et al. did,
the entropic repulsion may have caused an effective repulsion 
in the $x$-direction and eventually the coarsening in our case, too.
       \begin{figure}[btph]
       \centerline{
       \includegraphics[width=0.65\linewidth]{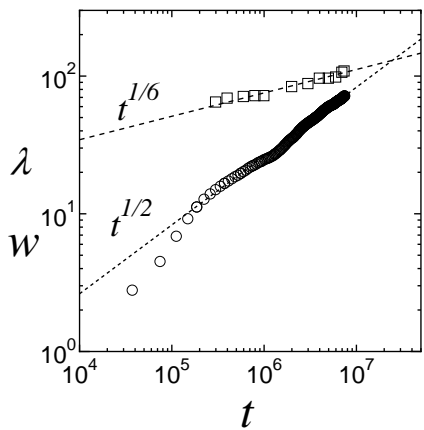}
       }
       \caption{
       Time evolution of the step width ($\bigcirc$), $w \sim t^{1/2}$,
       and the groove wavelength ($\Box$), $\lambda \sim t^{1/6}$.
       }\label{fig:width}
       \end{figure} 

In Figure~\ref{fig:width} open circles represent the time evolution
of the wandering amplitude $w$ for $f=-0.1$. 
The result is obtained by averaging over 10 runs
of the size $512 \times 128$ with 16 steps.
In an early stage ($t \le 2 \times 10^5$),
the step width increases rapidly.
Then, the width  enhancement slows down to $w \approx t^{1/2}$.
The exponent is the same with the values obtained for the step wandering 
in other conserved 
systems~\cite{Sato-ush2002prb,Pierre-Louis-mskp98prl,%
Gillet-pm00epjp,Paulin-gpm01prl,kato-ush}.
The slowing down of the fluctuation amplification 
is attributed to the suppression of the diffusion current due 
to the narrowing of the terrace width~\cite{Paulin-gpm01prl}. 
The groove wavelength 
$\lambda$
is obtained by counting the
number of grooves for 10 samples, 
and is depicted by open squares in Fig.~\ref{fig:width}.
The slow increase as $\lambda \sim t^{\alpha}$ with $\alpha=0.17 \pm 0.04$
is consistent with  the one found with the use of a generalized 
version of eq.(4)~\cite{Paulin-gpm01prl}.

In this paper,
we studied the drift-induced step wandering on a vicinal face
with  an anisotropic surface diffusion whose orientation dependence 
alternates on consecutive terraces.
The step-step interaction is shown to play an essential role 
for the step wandering, since it not only prevents steps from
coalescing but also creates the difference $\Delta c$
of the equilibrium adatom
density. The imbalance of the diffusion current between different steps 
induced by $\Delta c$ causes the step wandering. 
Thus the step repulsion mediates steps to establish the asymmetry 
in the diffusion field. Due to this asymmetry,
the step wandering occurs with step-up drift.
The in-phase wandering leads to the formation of straight grooves
on a vicinal face, in accordance with other conserved 
systems~\cite{Pierre-Louis-mskp98prl,Gillet-pm00epjp,Paulin-gpm01prl,
Sato-ush2002prb,kato-ush}.

Recently, in an experiment by Nielsen and coworkers~\cite{Nielsen-pp01ss}
a dimpled specimen was used and the step wandering on Si(001) vicinal face
was observed with the application of a direct electric current.
Near the bottom of the dimple, where the inclination is very small,
the step bunching occurs irrespective of the current direction~\cite{note}.
Fluctuation of bunches with the step-up current is larger
than that with the step-down current. On  increasing the inclination,
which means increasing the repulsive interaction,
the in-phase step wandering occurs and straight grooves parallel
to the current appear with the step-up current.
The step wandering was observed in a range of inclination angle between 
$0.08^\circ$ and $0.5^\circ$. In Si(001) vicinal face,
the direction of the drift of adatoms is believed to be the same 
as that of the electric current~\cite{Metois-hp99ss}, and
our results of step wandering under a step-up drift 
qualitatively agree  with the experiment~\cite{Nielsen-pp01ss}.

This work was  supported by Grant-in-Aid for Scientific Reserch
from Japan Society  of the Promotion of Science.
M.U. and Y.S. benefited from the inter-university
cooperative research program of the Institute for
Materials Research, Tohoku University.



\begin{references}
\bibitem{note}
In Ref.\cite{Nielsen-pp01ss} this phenomenon is used 
to determine the drift direction on anisotropic surfaces. 
However, we demonstrated in a Monte Carlo simulation that
 the bunching occurs irrespecteve of the current direction~\cite{Sato-us02jcg}.
\end{references}
\end{document}